\documentstyle[12pt,epsfig]{article}

\newcommand{\beq}{\begin{eqnarray}}
\newcommand{\eeq}{\end{eqnarray}}
\newcommand{\ba}{\begin{array}}
\newcommand{\ea}{\end{array}}

\newcommand{\nn}{\nonumber}

\newcommand{\raw}{\rightarrow}
\newcommand{\lsim}{\raisebox{-0.13cm}{~\shortstack{$<$ \\[-0.07cm] $\sim$}}~}

\def\plb#1#2#3{{\it Phys. Lett. }{\bf B#1~}(19#2)~#3}

\def\npb#1#2#3{{\it Nucl. Phys. }{\bf B#1~}(19#2)~#3}

\def\hepph#1{{\bf hep-ph}/#1}
\begin{document}
%
\renewcommand{\thefootnote}{\alph{footnote}}
\begin{titlepage}
\vspace*{-1cm}
\phantom{bla}
\hfill{hep-ph/9805313}
\vskip 2.0cm
\begin{center}
{\Large \bf One-loop Weak Dipole Form Factors and\\
\vskip 0.4truecm
Weak Dipole Moments of Heavy Fermions }
\end{center}
\vskip 1.5cm
\begin{center}
{\large Stefano Rigolin \footnote{Work partially supported by the EC under 
                             contract ERBFMBICT972474}} \\
\vskip .1cm
{{\it Departamento de F\`{\i}sica Te\`orica, Universidad Aut\`onoma de Madrid, 
SPAIN} \\
 {\rm E--Mail: rigolin@delta.ft.uam.es} }
\end{center}
\vskip 3.0truecm
%
\begin{center}
\bf Abstract
\end{center}
%
\begin{quote}
The one--loop weak--magnetic and weak--electric dipole form factors of
heavy fermions in a generic model are derived. Numerical predictions for 
the $\tau$ lepton and $b$ quark Weak Anomalous Magnetic and Electric Dipole 
Moments (AWMDM and WEDM) in the SM and MSSM are reviewed. The  MSSM 
contribution to the $\tau$ ($b$) AWMDM could be, in the high $\tan\beta$ 
scenario, four (thirty) times larger than the Electroweak SM 
one, but still a factor five below the QCD contribution (in the $b$ case). 
More interesting is the CP--odd sector where the contribution to the 
$\tau$ ($b$) WEDM in the MSSM could be up to twelve orders of magnitude 
larger than in the SM.  
\end{quote}
%
%
\vfill{Talk given at the XXXIIIrd Rencontres de Moriond}
\end{titlepage}
\setcounter{footnote}{0}
\renewcommand{\thefootnote}{\arabic{footnote}}
\baselineskip=18pt
%
\subsection*{$Vff$ effective vertex for on--shell fermions}
%
The most general $Vff$ effective vertex describing the interaction 
between a neutral vector boson and two on--shell fermions can be 
conventionally written in terms of six independent form factors as:  
\vskip -0.6truecm
\beq
\Gamma^{Vff}_\mu (s)=
   {\rm i} \Big\{\gamma_\mu \Big[F^V_{\rm V} - F^V_{\rm A} \gamma_5 \Big] - 
   (q+\bar{q})_\mu \Big[{\rm i} F^V_{\rm S} + F^V_{\rm P} \gamma_5 \Big] + 
   \sigma_{\mu\nu} (q+\bar{q})^\nu \Big[{\rm i} F^V_{\rm M} + F^V_{\rm E}
   \gamma_5 \Big] \Big\}. 
\label{vertex}
\eeq
Here $q$ and $\bar{q}$ are respectively the outgoing momenta of the fermion 
and the antifermion and $s$ is the square of the total momentum $p = q + 
\bar{q}$. The form factors $F^V_i$ are, in general, functions of the total 
energy $s$ and of all the other possible kinematic invariants of the process. 
$F^V_{\rm V}$ and $F^V_{\rm A}$ are the usual vector and axial--vector 
form factors. Being related to $D=4$ operators they are the only terms 
that can appear, at tree--level, in the Lagrangian of a renormalizable theory. 
$F^V_{\rm S}(s)$ and $F^V_{\rm P}(s)$ are the so--called 
scalar and pseudo--scalar form factors. They are usually negligible. 
Finally $F^V_{\rm M}(s)$ and $F^V_{\rm E}(s)$ are known as magnetic and 
electric form factors. The Anomalous Magnetic Dipole Moment (AMDM) 
and the Electric Dipole Moment (EDM), associated to a neutral vector boson 
$V$, are defined as:
\vskip -0.6truecm
\beq
      a^V_f = \frac{2 m_f}{e} F^V_{\rm M}(s=M^2_V) \qquad & {\rm and} & 
      \qquad d^V_f = - F^V_{\rm E}(s=M^2_V). 
\label{defwdm}
\eeq
Here $e$ is the electron charge, $m_f$ and $M_V$ are the fermion and 
boson masses. If $V=\gamma$ ($M_\gamma=0$) Eq.~\ref{defwdm} reproduces  
the usual definitions of the photon AMDM and EDM. For $V=Z$ Eq.~\ref{defwdm} 
defines the Anomalous Weak Magnetic Dipole Moment (AWMDM = $a^w_f$) and 
the Weak Electric Dipole Moment (WEDM = $d^w_f$). Although the formulation 
could be completely general in the following we concentrate our analysis on 
the Weak Dipole Form Factors (WDFFs). 
%
%
%
%
\subsection*{One--loop generic expressions of the Weak Dipole Form Factors}
\begin{figure}[t]
\begin{center}
\vskip -1truecm
\epsfig{figure=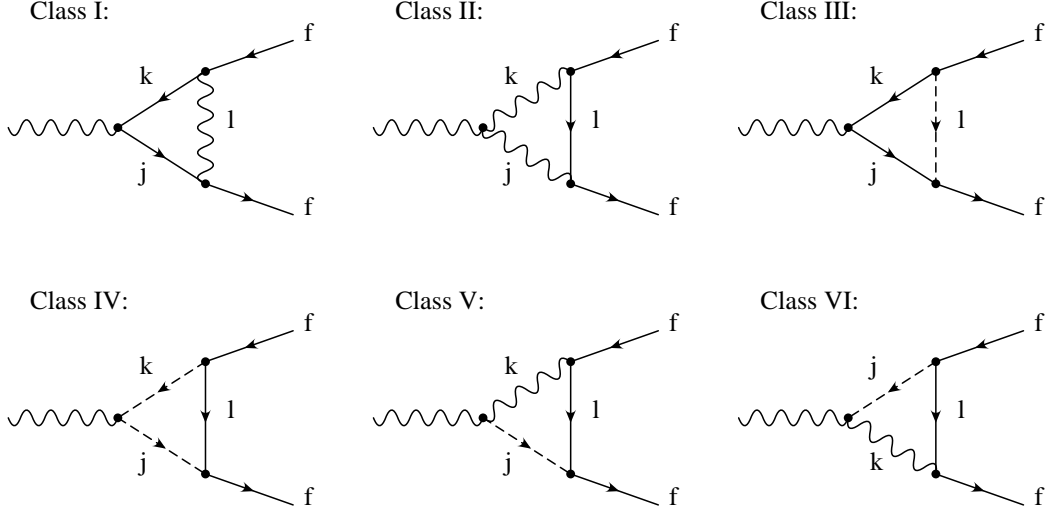,height=7cm,width=14cm,angle=0}
\end{center}
\caption{One-loop topologies for a general $Vff$ effective vertex.}
\label{fig1}
\vskip -0.5truecm
\end{figure}

%
All the possible one--loop contributions to the $a^Z_f(s)$ and $d^Z_f(s)$ 
form factors can be classified in terms of the six classes of triangle 
diagrams depicted in Fig.~(\ref{fig1}). 
The vertices are labelled by generic couplings, according to the following 
interaction Lagrangian, for vector bosons $V^{(k)}_\mu=A_\mu,\ Z_\mu,\ W_\mu,
\ W^\dagger_\mu$, fermions $\Psi_k$ and scalar bosons $\Phi_k$:
\vskip -0.6truecm
\beq
{\cal L} 
& = & 
     {\rm i} e J (W^\dagger_{\mu\nu}W^\mu Z^\nu - W^{\mu\nu}W^\dagger_\mu Z_\nu
             + Z^{\mu\nu}W^\dagger_\mu W_\nu)
     + e V^{(k)}_\mu\bar{\Psi}_j\gamma^\mu(V^{(k)}_{jl}-A^{(k)}_{jl}\gamma_5)
       \Psi_l \nn \\
& & 
     + \Big\{e \bar{\Psi}_f(S_{jk}-P_{jk}\gamma_5)\Psi_k\Phi_j\ 
     + e K_{jk} Z^\mu V^{(k)}_\mu\Phi_j + {\rm h.c.} \Big\} 
     + {\rm i} e G_{jk} Z^\mu\Phi_j^\dagger\stackrel{\leftrightarrow}
             {\partial}_\mu\Phi_k.
\label{genlag}
\eeq
Every class of diagrams is calculated analytically and expressed in terms 
of the couplings introduced in (\ref{genlag}) and the one--loop three--point 
integrals. The computation is done in the 't Hooft-Feynman gauge. 
Similar expressions are also derived in Ref.[1,2,3].
%
%
\newline
\noindent
-- [Class I]: vector boson exchange contribution.
\vskip -0.6truecm
\beq
\frac{a^Z_f}{2 m_f}({\rm I}) = \frac{\alpha}{4\pi} \Big\{ 
             4\sum_{jkl} m_k {\rm Re} 
     \Big[V^{(Z)}_{jk}(V^{(l)}_{fj}V^{(l)*}_{fk} - 
    A^{(l)}_{fj}A^{(l)*}_{fk}) - A^{(Z)}_{jk}(V^{(l)}_{fj}A^{(l)*}_{fk} - 
    A^{(l)}_{fj}V^{(l)*}_{fk})\Big] \Big[2C^{+}_1-C_0\Big]_{kjl} \nn
\eeq
\vskip -0.6truecm
\beq
            +4m_f\sum_{jkl}{\rm Re} \Big[
    V^{(Z)}_{jk}(V^{(l)}_{fj}V^{(l)*}_{fk}+A^{(l)}_{fj}A^{(l)*}_{fk}) + 
    A^{(Z)}_{jk}(V^{(l)}_{fj}A^{(l)*}_{fk}+A^{(l)}_{fj}V^{(l)*}_{fk})\Big] 
        \Big[2C^+_2 - 3C^+_1 + C_0\Big]_{kjl} \Big\},
\label{mdm1}
\eeq
\vskip -0.6truecm
\beq
\frac{d^Z_f}{e}({\rm I}) = \frac{\alpha}{4\pi}  \Big\{ 
            4m_f\sum_{jkl}{\rm Im}
    \Big[V^Z_{jk}(V^{(l)}_{fj}A^{(l)*}_{fk}+A^{(l)}_{fj}V^{(l)*}_{fk}) + 
     A^{(Z)}_{jk}(V^{(l)}_{fj}V^{(l)*}_{fk}+A^{(l)}_{fj}A^{(l)*}_{fk})\Big] 
    \Big[2C^{\pm}_2-C^-_1 \Big]_{kjl} \nn
\eeq
\vskip -0.6truecm
\beq
           -4\sum_{jkl}m_k{\rm Im}
    \Big[V^{(Z)}_{jk}(V^{(l)}_{fj}A^{(l)*}_{fk}-A^{(l)}_{fj}V^{(l)*}_{fk})-
    A^{(Z)}_{jk}(V^{(l)}_{fj}V^{(l)*}_{fk}-A^{(l)}_{fj}A^{(l)*}_{fk})\Big]
    \Big[2C^{+}_1-C_0 \Big]_{kjl} \Big\}.
\label{edm1}
\eeq
A particular but relevant subcase of Class I is the gluon exchange 
contribution. The derivation of the corresponding formula is straightforward 
once one performs the substitutions $\alpha \raw \alpha_S$, 
$V^{(l)}_{fi} \raw T_l$ (the $SU(3)$ generators) and $A^{(l)}_{fj} \raw0$. 
\newline
\noindent 
-- [Class II]: fermion exchange contribution with two internal vector bosons.
\vskip -0.6truecm
\beq
\frac{a^Z_f}{2 m_f}({\rm II}) 
&=& 
       \frac{\alpha}{4\pi} \Big\{ 2m_f\sum_{jkl}{\rm Re}
   \Big[J (V^{(j)}_{fl}V^{(k)*}_{fl} + A^{(j)}_{fl}A^{(k)*}_{fl}) \Big] 
             \Big[4C^{+}_2 + C^+_1 \Big]_{kjl} \nn \\
& &
   \hspace{1truecm} - 6\sum_{jkl} m_l{\rm Re} 
   \Big[J(V^{(j)}_{fl}V^{(k)*}_{fl} - A^{(j)}_{fl}A^{(k)*}_{fl}) \Big]
             \Big[C^+_1 \Big]_{kjl} \Big\}, \\
\label{mdm2}
\frac{d^Z_f}{e}({\rm II}) 
&=&
       -\frac{\alpha}{4\pi} \Big\{ 2m_f\sum_{jkl}{\rm Im}
   \Big[J(V^{(j)}_{fl}A^{(k)*}_{fl}+A^{(j)}_{fl}V^{(k)*}_{fl}) \Big]
            \Big[4C^{\pm}_2-C^-_1 \Big]_{kjl} \nn \\
& &
  \hspace{1truecm} + 6\sum_{jkl} m_l{\rm Re}
   \Big[J(V^{(j)}_{fl}A^{(k)*}_{fl}-A^{(j)}_{fl}V^{(k)*}_{fl}) \Big]
           \Big[ C^+_1 \Big]_{kjl} \Big\}.
\label{edm2}
\eeq
-- [Class III]: scalar boson exchange contribution.
\vskip -0.6truecm
\beq
\frac{a^Z_f}{2 m_f}({\rm III}) &=&
       \frac{\alpha}{4\pi}  \Big\{2m_f\sum_{jkl}{\rm Re}
   \Big[V^{(Z)}_{jk}(S_{lj}S^*_{lk} + P_{lj}P^*_{lk}) +
   A^{(Z)}_{jk}(S_{lj}P^*_{lk} + P_{lj}S^*_{lk}) \Big] 
   \Big[2C^{+}_2 - C^+_1 \Big]_{kjl} \nn \\
& & 
         \hspace{-0.5truecm}
   - 2\sum_{jkl} {m}_k {\rm Re}\Big[V^{(Z)}_{jk}(S_{lj}S^*_{lk}-P_{lj}P^*_{lk})
   -A^{(Z)}_{jk}(S_{lj}P^*_{lk}-P_{lj}S^*_{lk}) \Big] 
   \Big[C^+_1 + C^-_1 \Big]_{kjl} \Big\}, \\
\label{mdm3}
\frac{d^Z_f}{e}({\rm III})
&=&  
         \frac{\alpha}{4\pi} \Big\{-2m_f\sum_{jkl}{\rm Im}
   \Big[V^{(Z)}_{jk}(P_{lj}S^*_{lk}+S_{lj}P^*_{lk})
   +A^{(Z)}_{jk}(S_{lj}S^*_{lk}+P_{lj}P^*_{lk}) \Big] 
   \Big[2C^{\pm}_2-C^-_1 \Big]_{kjl} \nn \\
& & 
         \hspace{-0.5truecm}
   +2\sum_{jkl} {m}_k {\rm Im}\Big[V^{(Z)}_{jk}(P_{lj}S^*_{lk}-S_{lj}P^*_{lk})
   +A^{(Z)}_{jk}(S_{lj}S^*_{lk}-P_{lj}P^*_{lk}) \Big] 
   \Big[C^+_1+C^-_1 \Big]_{kjl} \Big\}.
\label{edm3}
\eeq
-- [Class IV]: fermion exchange contribution with two internal scalar bosons.
\vskip -0.6truecm
\beq
\frac{a^Z_f}{2 m_f}({\rm IV})
& = & 
       - \frac{\alpha}{4\pi}  \Big\{ 2 m_f\sum_{jkl}{\rm Re}
   \Big[G_{jk}(S_{jl}S^*_{kl} + P_{jl}P^*_{kl}) \Big]
   \Big[2C^{+}_2 - C^+_1 \Big]_{kjl} \nn \\
& & \hspace{0.8truecm}
   +\sum_{jkl} {m}_l {\rm Re}\Big[G_{jk}(S_{jl}S^*_{kl} - P_{jl}P^*_{kl}) \Big] 
   \Big[2C^+_1 - C_0 \Big]_{kjl} \Big\}, \\
\label{mdm4}
\frac{d^Z_f}{e}({\rm IV})
&=&
         \frac{\alpha}{4\pi}  \Big\{2m_f\sum_{jkl} {\rm Im}
  \Big[G_{jk}(S_{jl}P^*_{kl}+P_{jl}S^*_{kl})\Big] 
  \Big[2C^{\pm}_2-C^-_1\Big]_{kjl} \nn \\  
& & \hspace{0.8truecm}
   -\sum_{jkl} {m}_l {\rm Im}\Big[G_{jk}(S_{jl}P^*_{kl} - P_{jl}S^*_{kl}) \Big]
   \Big[2C^+_1 - C_0 \Big]_{kjl} \Big\}.
\label{edm4}
\eeq
-- [Class V+VI]: fermion exchange contribution with one vector and one 
                    scalar internal boson.
\vskip -0.6truecm
\beq
\frac{a^Z_f}{2 m_f}({\rm V+VI})
&=&
         \frac{\alpha}{4\pi} 2\sum_{jkl}{\rm Re} 
   \Big[K_{jk}(V^{(k)}_{fl}S_{jl}^*+A^{(k)}_{fl}P_{jl}^*) \Big]
   \Big[C^+_1+C^-_1 \Big]_{kjl}, \\ 
\label{mdm56}
\frac{d^Z_f}{e}({\rm V+VI})
&=&
         -\frac{\alpha}{4\pi} 2\sum_{jkl}{\rm Im}
   \Big[K_{jk}(V^{(k)}_{fl}P_{jl}^*+A^{(k)}_{fl}S_{jl}^*) \Big]
   \Big[C^+_1+C^-_1 \Big]_{kjl}.
\label{edm56}
\eeq
In the Eqs.(\ref{mdm1}--\ref{edm56}) the shorthand notation $[C]_{kjl}$ 
stands for $C(-\bar{q},q,M_k,M_j,M_l)$. The definition of the $C$ integrals 
used and the relations with the conventional set of three-point function 
integrals can be found in [4]. 
All the expressions (\ref{mdm1}--\ref{edm56}) are, at least, proportional 
to a fermion mass (either internal or external), consistently with the 
chirality flipping character of the dipole moments. In class V and VI 
diagrams the fermion mass proportionality arises in the product of the 
Yukawa couplings $S$ and $P$ and the mass-dimension term $K$. 
From this fact follows that heavy fermions seem, in general, to be the best 
candidates for an experimental analysis. Hence, for on-shell $Z$ bosons, 
the $b$ quark and $\tau$ lepton are the most promising options. 
Eqs.(\ref{mdm1}--\ref{edm56}) show also that, in general, the DMs for 
massless fermion are not vanishing but proportional to final fermion 
masses running in the loop. The SM cancellation of the massless neutrino DMs 
is only ``accidental''. Finally notice that all the contributions to 
the WEFFs are proportional to the imaginary part of a certain combination of 
couplings. A theory with only real couplings has, manifestly, 
vanishing (W)EFFs.
%
\subsection*{The $\tau$ and $b$ WDMs in the SM}
%
Our numerical evaluation of the SM $\tau$ and $b$ AWMDM are in perfect 
agreement with Ref.~[5]. Taking as input parameters 
$m_{\tau}=1.777$ GeV, $m_b=4.5$ GeV, $m_t=175$ GeV, $M_Z=91.19$ GeV, 
$s^2_W=0.232$, $\alpha=1/128$ and $\alpha_s=0.118$, the pure electroweak 
contribution are $a^w_{\tau}({\rm EW}) = (2.10 + 0.61~{\rm i})\times 10^{-6}$
and $a^w_b({\rm EW}) = ([1.1;\ 2.0;\ 2.4] - 0.2~{\rm i})\times10^{-6}$ 
for three different values of the Higgs mass (respectively $M_{H^0}=M_Z$, 
$2M_Z$ and $3M_Z$). 
$V_{tb}$ equal to one and off-diagonal entries equal to zero are assumed 
in the CKM matrix. But also QCD affects the $Zb\bar{b}$ vertex at one loop. 
The gluon exchange dominates being almost a factor one hundred larger than 
the weak contributions. The whole SM $b$ AWMDM is then  
$a^w_b[{\rm SM}] = (-2.98 + 1.56~{\rm i}) \times 10^{-4}$. 

The only phase present in the SM, the $\delta_{CKM}$, it is not sufficient 
for generating one-loop contributions to the (W)EDM\footnote{We implicitly 
assume here vanishing $\theta_{QCD}$ phase.}. It is also proved that such 
CP violating terms cannot appear even at two-loop level. A very crude 
estimate of the three-loop SM contribution, using simple power counting 
arguments, gives the indicative limits for the $\tau$ and $b$ (W)EDM:  
$d^w_{\tau} / \mu_\tau \leq 1.7\times 10^{-19}$ and $d^w_b / \mu_b \leq 
1.4\times 10^{-18}$. 
The ``magnetons'' $\mu_\tau = 1.7 \times 10^{-15}$ and $\mu_b =0.7 
\times 10^{-15}$ $e$cm are useful for rendering the (W)EDMs dimensionless. 
%
\subsection*{The $\tau$ and $b$ WDMs in the MSSM}
%
The particle content of the MSSM comes from the SM spectrum with two 
substantial modifications: the enlargement of the Higgs sector from one 
to two doublets and the appearance of all the SUSY partners of the standard 
particles. In a R--parity conserving formulation the sets of new ``genuine'' 
MSSM diagrams are: i) diagrams with the MSSM two-doublet Higgs sector;
ii) diagrams with charginos (and sfermions); iii) diagrams with neutralinos 
(and sfermions); iv) diagrams with gluinos (and sfermion) in the $b$ case.

The MSSM contribution to the imaginary part of the $\tau$ or $b$ AWMDM is 
generally negligible. Only strong MSSM threshold effects (typically 
originated by light neutralinos ) can produce contributions comparable 
to the SM ones. 

The real part of the $\tau$ AWMDM is dominated by the chargino contribution 
and is roughly proportional to $\tan\beta$. Neutralino and  Higgs sector 
contributions are negligible in most of the experimentally allowed MSSM 
parameter space [1].  The total contribution can reach the value of $|{\rm Re} 
(a^w_\tau[{\rm MSSM}])| \sim 0.5~(8) \times 10^{-6}$ for $\tan\beta = 1.6~(50)$, 
so being, in the most favourable case, four times larger than in the SM. 

The most important MSSM contributions to the real part of the $b$ AWMDM 
is provided by charginos and gluinos in the high $\tan\beta$ scenario.
The total contribution to the real part can reach the value of 
$|{\rm Re}(a^w_b[{\rm MSSM}])| \sim 2~(50) \times 10^{-6}$ for $\tan\beta = 
1.6~(50)$. The high $\tan\beta$ values are one order of magnitude higher 
than the pure electroweak SM contribution, but still a factor five below 
the standard QCD contribution [1].

A full scan of the MSSM parameter space has been performed in search for 
the maximum effect on the WEDM of the $\tau$ lepton and the $b$ quark [3]. 
The Higgs sector does not contribute and chargino diagrams are more 
important than neutralino ones. Gluinos are also involved in the $b$ case 
and compete in importance with charginos. In the most favourable
configuration of CP-violating phases and for values of the rest of the
parameters still not excluded by experiments, these WEDMs can be as much
as twelve orders of magnitude larger than the SM predictions 
$|{\rm Re} ( d^w_\tau )| \lsim 0.2~(6) \times 10^{-6}~\mu_\tau\quad {\rm and} 
\quad |{\rm Re} ( d^w_b )| \lsim 2~(35) \times 10^{-6}~\mu_b$.
There may be a contribution to the imaginary part if the neutralinos are 
light but this contribution is at least one order of magnitude less 
than the real part of the $\tau$ or bottom WEDM.
%
\subsection*{Acknowledgements}
%
I'm indebted to W.~Hollik, J.I.~Illana and D.~St\"ockinger for the very 
pleasant collaboration on which this talk is based. I would like 
to thank the organizers and all the participants at the ``XXXIIIrd Rencontres 
de Moriond'' for the very nice atmosphere enjoyed in Les Arcs. 
%
\subsection*{References}
%
\begin{itemize}
\item[{[1]}]
        W. Hollik, J. I. Illana, S. Rigolin and D. St\"ockinger, 
        \plb{416}{98}{345};
        B. de Carlos, J.M. Moreno,
        \hepph{9707487}.
\item[{[2]}]
        A. Bartl, E. Christova, W. Majerotto, 
        \npb{460}{96}{235} [E: {\em ibid.} {\bf B465} (1996) 365];  
        A. Bartl, E. Christova, T. Gajdosik, W. Majerotto, \npb{507}{97}{35}.
\item[{[3]}] 
        W. Hollik, J. I. Illana, S. Rigolin and D. St\"ockinger, 
        \hepph{9711322} (to be published in {\em Phys. Lett.} {\bf B}).
\item[{[4]}]
        W. Beenakker, S.C. van der Marck and W. Hollik, \npb{365}{91}{24}.
\item[{[5]}]
	{J. Bernab\'eu, G.A. Gonz\'alez-Sprinberg, J. Vidal},
	\plb{326}{94}{168} and {\em ibid.} {\bf B397} (1997) 255;
	{J. Bernab\'eu, G.A. Gonz\'alez-Sprinberg, M.Tung, J. Vidal},
	\npb{436}{95}{474}.
\end{itemize}
%
\end{document}